\documentclass[runningheads]{llncs}
\usepackage[T1]{fontenc}
\usepackage{amsmath}
\usepackage{multicol}
\usepackage{amsfonts}
\usepackage{xcolor}
\usepackage{graphicx}
\usepackage{tcolorbox}

\usepackage{tcolorbox}
\usepackage{tikz}
\usepackage[linesnumbered, ruled, vlined]{algorithm2e}
\usepackage{caption}
\usepackage{subcaption}
\usepackage{cite}

\usepackage{url}
\usepackage{hyperref}
\hypersetup{
    colorlinks=true,
    linkcolor=blue,
    filecolor=magenta,      
    urlcolor=cyan,
}
\begin{document}
\title{Balanced Popularity in Multi-Product Billboard Advertisement}
\author{Dildar Ali \and Suman Banerjee \and Yamuna Prasad}
\authorrunning{Ali et al.} 
\institute{Department of Computer Science and Engineering, Indian Institute of Technology Jammu, Jammu \& Kashmir-181221, India. \\
\email{\{2021rcs2009,suman.banerjee,yamuna.prasad\}@iitjammu.ac.in}}
\maketitle
\begin{abstract}
 The billboard advertisement has emerged as an effective out-of-home advertisement technique where the objective is to choose a limited number of slots to play some advertisement content (e.g., animation, video, etc.) with the hope that the content will be visible to a large number of travelers, and this will be helpful to earn more revenue. In this paper, we study a variant of the influential slot selection problem where the advertiser wants to promote multiple products.  Formally, we call this problem the \textsc{Multi-Product Influence Maximization Problem for the Balanced Popularity} Problem. The input to our problem is a trajectory and a billboard database, as well as a budget for each product. The goal here is to choose a subset of slots for each product such that the aggregated influence of all the products gets maximized subject to the following two constraints: total selection cost for each product is less than or equal to the allocated budget for that product, and the difference between the influence for any two products is less than or equal to a given threshold. We show that the problem is NP-hard to solve optimally. We formulate this problem as a linear programming problem and use linear programming relaxation with randomized rounding. Further, we propose a greedy-based heuristic with balance correction to solve this problem. We conduct a number of experiments with real-world trajectory and billboard datasets, and the results are reported. From the reported results, we observe that the proposed solution approaches lead to more influence compared to many baseline methods.  
\end{abstract}
\section{Introduction} Recently, \emph{Billboard Advertisement} has emerged as an effective out-of-home advertising technique due to the guaranteed return on investment\footnote{\url{https://www.lamar.com/howtoadvertise/Research/}}.
In a billboard advertisement, a number of digital billboards are placed across popular places in a city, and the advertisement contents (e.g., animation, video, etc.) are played with the hope that the content will be observed by many people. This is a popular out-of-home advertisement technique that is helpful to reach a large audience with a low budget to create influence and awareness of the brand. Eventually, a little investment in advertising may lead to a large revenue with high probability. 



\paragraph{\textbf{Background.}} To our knowledge, Zhang et al. \cite{zhang2018trajectory,zhang2020towards} are the first to introduce the influence maximization problem in billboard advertising. This problem has been studied in the literature, and a number of solution methodologies have been proposed, such as the graph-based pruned-submodularity approach \cite{ali2022influential}, the spatial clustering approach \cite{ali2023influential}, the branch-and-bound approach \cite{ali2024influentialzonal,zhang2019optimizing}, and many more. Wang et al. \cite{wang2019efficiently} introduce the problem of finding the $k$ best advertising units in billboard advertisements. There also exists literature in the context of advertising on billboards \cite{ali2024effective,ali2024influential} and social networks \cite{ke2018finding} that select tags and slots jointly. Deviating from the influence maximization, there exists literature \cite{zhang2021minimizing,ali2024minimizing,aslay2015viral} that considers regret minimization in billboard advertising. In practice, an e-commerce house has multiple products to advertise. However, to our knowledge, there is no literature that considers slot selection for multiple products. Hence, in this paper, we address the \textsc{Multi-Product Influence Maximization Problem for the Balanced Popularity} Problem. 

\paragraph{\textbf{Motivation.}}
Consider the scenario where a commercial house wants to promote more than one product through billboard advertising. Different products will have different features, and hence it may not be meaningful to promote all the products to the same set of people. To make the campaign more effective, it is important to select slots that cover the people who are more likely to like the product. In this paper, we study the problem of selecting slots for multiple products where the goal is to maximize the aggregated influence for all products. However, it is important to observe that it may not be beneficial for the commercial house if there is a significant difference in influence between any two products. Because if for any product its influence is very low compared to another product, then the expected sales of the product will be low, and the initial investment of the commercial house for launching the product will go in vain. As a remedy, the commercial house may wish to choose slots in such a way as to lead to a balanced influence for all the products.  


\paragraph{\textbf{Our Contribution.}} To the best of our knowledge, this problem has not been studied previously, and we make the following contributions.
\begin{itemize}
\item We study the noble problem called the \textsc{Multi-Product Influence Maximization Problem for the Balanced Popularity} problem for which there is no literature.
\item We formulate this problem as a linear programming problem and
use linear programming relaxation with randomized rounding. Further,
we propose a greedy-based heuristic with balance correction to solve this
problem.
\item  We conducted experiments with real-world trajectory and billboard datasets to show the effectiveness and efficiency of the proposed solution approach.
\end{itemize}

\paragraph{\textbf{Organization of the Paper}} The rest of the paper has been organized as follows. Section \ref{Sec:Problem} describes the background information and defines the problem formally. The proposed solution approaches have been described in Section \ref{Sec:Solution}. Section \ref{Sec:Experiments} describes the experimental evaluation of the proposed solution approaches. Section \ref{Sec:Conclusion} describes the concluding remarks of our study.

\section{Background and Problem Definition} \label{Sec:Problem}
\paragraph{\textbf{Trajectory and Billboard Database.}}
A trajectory database contains the location information of a set of people in a city across different time stamps. In our problem, a trajectory database $\mathcal{D}$ contains tuples of the form $(u_i, \texttt{loc},[t_a,t_b], \mathcal{P}(u_i))$ and this signifies that the person $u_i$ is at the location $\texttt{loc}$ for the duration $t_a$ to $t_b$. Here, $\mathcal{P}(u_i)$ denotes the set of products in which user $u_i$ will be interested. For any person $u_i$, $loc_{[t_x,t_y]}(u_i)$ denotes the locations of $u_i$ during the time interval $[t_x,t_y]$. Let $\mathcal{U}$ denote the set of people whose movement data is contained in $\mathcal{D}$. Let, $T_{1}= \underset{\tau \in \mathcal{D}}{min} \ t_a$ and $T_{2}= \underset{t \in \mathcal{D}}{max} \ t_b$ and we say that the trajectory database $\mathcal{D}$ contains the movement data from time stamp $T_1$ to $T_2$. A billboard database $\mathbb{B}$ contains the information about billboard slots. Typically, this contains the tuples of the following form $(b_{id}, s_{id}, \texttt{loc}, \texttt{duration})$ where $b_{id}$ and $s_{id}$ denote the billboard id and slot id. The \texttt{loc} and \texttt{duration} signify the billboard's location and the slot's duration.
\paragraph{\textbf{Billboard Advertisement.}}
Let a set of $m$ billboards $\mathcal{B}=\{b_1, b_2, \ldots, b_m\}$ be placed in different locations of a city. We assume that all the billboards can be leased for a multiple of a fixed duration $\Delta$, which is called a `slot' and has been stated in Definition \ref{Def:Slot}.
\begin{definition} [Billboard Slot] \label{Def:Slot}
A billboard slot is defined by a tuple consisting of two entities, the billboard ID and the duration, that is, $(b_{id}, [t,t+\Delta])$. 
\end{definition}
The set of all billboard slots is denoted by $\mathcal{BS}$, i.e., $\mathcal{BS}=\{(b_i, [t,t+\Delta]): i \in \{1,2, \ldots, m\} \text{ and } t \in \{T_1, T_1+\Delta, T_1+ 2 \Delta, \ldots, T_2-\Delta \}\}$. It can be observed that $|\mathcal{BS}|=m \cdot \frac{T}{\Delta}$ where $T=T_{2}-T_{1}$. For simplicity, we assume that $T$ is perfectly divisible by $\Delta$. For any slot $s_j \in \mathcal{BS}$, $b_{s_j}$ denotes the corresponding billboard and $[t^{s}_{s_j}, t^{f}_{s_j}]$ denotes the time interval for the slot $s_j$ where $t^{f}_{s_j}=t^{s}_{s_j} + \Delta$. Now, we state the notion of influence probability of a billboard slot in Definition \ref{Def:Inf_Prob}.
\begin{definition} [Influence Probability of a Billboard slot]  \label{Def:Inf_Prob}
Given a slot $s_j \in \mathcal{BS}$ and a person $u_i \in \mathcal{U}$, the influence probability of $s_j$ on $u_i$ is denoted by $Pr(s_j, u_i)$ and can be computed using the following conditional equation.
\[
    Pr(s_j,u_i)= 
\begin{cases}
    \frac{size(s_j)}{\underset{s_k \in \mathcal{BS}}{max} \ size(b_{s_k})},& \text{if } loc_{[t^{s}_{s_j}, t^{f}_{s_j}]}(u_i)=loc(s_j) \\
    0,              & \text{otherwise}
\end{cases}
\]
\end{definition}
Now, we define the influence of a subset of billboard slots in \ref{Def:Influence}.
\begin{definition} [Influence of Billboard Slots] \label{Def:Influence}
Given a trajectory database $\mathcal{D}$, and a subset of billboard slots $\mathcal{S} \subseteq \mathcal{BS}$, the influence of $\mathcal{S}$ can be defined as the expected number of trajectories that are influenced, which can be computed using Equation \ref{Eq:Equation}.
\begin{equation} \label{Eq:Equation}
\mathcal{I}^{'}(\mathcal{S})= \underset{u_i \in \mathcal{U}}{\sum} [1- \underset{s_j \in \mathcal{S}}{\prod} (1-Pr(s_j,u_i))]
\end{equation}
\end{definition}
Here, $Pr(b_j,u_i)$ denotes the influence probability of the billboard slot $b_j$ on the people $u_i$, and $\mathcal{I}^{'}(\mathcal{S})$ denotes the influence of the billboard slots of $\mathcal{S}$. $\mathcal{I}^{'}()$ is the influence function which maps each possible subset of billboard slots to its corresponding influence value, i.e., $\mathcal{I}^{'}: 2^{\mathcal{BS}} \longrightarrow \mathbb{R}_{0}^{+}$ where $\mathcal{I}^{'}(\emptyset)=0$. 
\par Now, it is important to observe that every product may not be relevant to every trajectory user. The effective advertisement requires targeting the right users. Let $\mathcal{P}= \{p_1,p_2,\ldots,p_{\ell}\}$ be the set of products. From the trajectory database, each user’s relevant products are known. For any product $j \in [p]$, we define Product Specific Influence for a billboard slot in Definition \ref{Def:Product_Influence}.

\begin{definition}[Product Specific Influence] \label{Def:Product_Influence}
 Given a specific product and a subset of billboard slots $\mathcal{S} \subseteq \mathcal{BS}$, the product-specific influence of $\mathcal{S}$ for the product $j$ is denoted by $\mathcal{I}_{j}(\mathcal{S})$ and can be computed using Equation \ref{Eq:product_influence}.
\begin{equation} \label{Eq:product_influence}
    \mathcal{I}_{j}(\mathcal{S})=  \underset{u_i \in \mathcal{U}_{k}}{\sum} [1- \underset{s_j \in \mathcal{S}}{\prod} (1-Pr(s_j,u_i))]
\end{equation}
Here, $\mathcal{U}_{k}$ denotes the set of users relevant to the $k$-th product and $\mathcal{I}$ is the product specific influence function, i.e., $\mathcal{I}: 2^{\mathcal{BS}} \times \mathcal{P} \longrightarrow \mathbb{R}^{+}_{0}$. 
\end{definition}
\paragraph{\textbf{Problem Definition.}}
Consider that a commercial company wants to promote $\ell$ many products through a billboard advertisement. The input to our problem is a trajectory and billboard database denoted by $\mathcal{D}$ and $\mathbb{B}$, respectively, $\ell$ positive integers $k_1, k_2, \ldots, k_{\ell}$ and a threshold value $\theta$. The goal of this problem is to find an allocation $(\mathcal{S}_1, \mathcal{S}_2, \ldots, \mathcal{S}_{\ell})$ where $\mathcal{S}_i$ denotes the allocated slots for the $i$-th product for all $i \in [\ell]$. Here, the objective is to maximize the sum of the influences of all products, i.e., to maximize $\underset{i \in [\ell]}{\sum} I_{i}(\mathcal{S}_i)$. The constraints involved in this problem are as follows:
\begin{itemize}
    \item \textbf{Budget Constraint}: This constraint ensures that for any product the number of allocated slots should be less than the corresponding budget, i.e., for all $i \in [\ell]$, $|\mathcal{S}_i| \leq k_i$.
    \item  \textbf{Non-Intersecting Slot Constraint}: This constraint ensures that for any $i,j \in [\ell]$ and $i \neq j$, $\mathcal{S}_i \cap \mathcal{S}_j =\emptyset$.
    \item \textbf{Influence Difference Constraint}: This constraint ensures that the influence between any two products will be less than equal to a given threshold, i.e., for all $i,j \in [\ell]$, and $i \neq j$, $|I_i(\mathcal{S}_i) - I_j(\mathcal{S}_j)| \leq \theta$ where $\theta$ is a given threshold value.
\end{itemize}
From the computational point of view, this problem can be posed as follows: 
\begin{tcolorbox}
\underline{\textsc{Multi-Product Influence Maximization Problem (MPIM)}} \\
\textbf{Input:} A trajectory ($\mathcal{D}$) and Billboard ($\mathcal{B}$) Database, A set of billboard slots ($\mathcal{BS}$), a set of products $(\mathcal{P})$.

\textbf{Problem:} Find a subset of slots for each product such that the aggregated influence of all the products gets maximized.

\end{tcolorbox}
By a reduction from the Set Cover Problem \cite{balas1972set}, we can show that the{Multi-Product Influence Maximization Problem is NP-hard. This result has been presented in Theorem \ref{NP-hard}. Due to the space limitation, we are not able to give the whole reduction. 
\begin{theorem}\label{NP-hard} The \textsc{Multi-Product Influence Maximization Problem} is NP-hard and hard to approximate in a constant factor.
\end{theorem}
\section{Proposed Solution Approaches} \label{Sec:Solution}
\subsection{\textbf{Linear Programming (LP) Relaxation}}
Let $\mathcal{BS}$ and $\mathcal{P}$ be the set of billboard slots and the set of products,  $\mathcal{U}_i\subseteq \mathcal{U}$ be the relevant users, and $k_i$ be the budget for the product $i$. $p_{s,u}\in[0,1]$ be the influence probability of the slot $s$ on user $u$.

\paragraph{\textbf{Decision variables.}} Let $x_{s,i}\in\{0,1\}$, where the slot $s$ is assigned to product $i$ and $y_{u,i}\in[0,1]$ be the auxiliary variable approximating the influence of product $i$ on user $u$.  The integer linear programming (ILP) formulation is stated below.

\paragraph{\textbf{Objective (linear surrogate of expected influence).}}
\begin{equation}
    \max \sum_{i\in \mathcal{P}}\ \sum_{u\in \mathcal{U}_i} y_{u,i}.
\end{equation}
\paragraph{\textbf{Constraints.}}
$$
\begin{aligned}
&\text{(Budget)} && \sum_{s\in \mathcal{BS}} x_{s,i} \le k_i && \forall~ i\in \mathcal{P},\\
&\text{(Disjointness)} && \sum_{i\in \mathcal{P}} x_{s,i} \le 1 && \forall~ s\in \mathcal{BS},\\
&\text{(Linking / coverage upper bound)} && y_{u,i} \le \sum_{s\in \mathcal{BS}} p_{s,u}\, x_{s,i} && \forall ~i\in \mathcal{P},\ \forall ~u\in \mathcal{U}_i,\\
& && 0 \le y_{u,i} \le 1 && \forall~ i\in \mathcal{P},\ \forall~ u\in \mathcal{U}_i,\\
&\text{(Balance)} && \sum_{u\in \mathcal{U}_i} y_{u,i} - \sum_{u\in \mathcal{U}_j} y_{u,j} \le \theta && \forall~ i\neq j,\\
& && \sum_{u\in \mathcal{U}_j} y_{u,j} - \sum_{u\in \mathcal{U}_i} y_{u,i} \le \theta && \forall~ i\neq j,\\
&\text{(Integrality)} && x_{s,i}\in\{0,1\} && \forall ~s\in \mathcal{BS},\ \forall~ i\in \mathcal{P}.
\end{aligned}
$$

\emph{Remark.} The constraint $y_{u,i} \leq \sum_s p_{s,u}x_{s,i}$ is the linear upper bound for the nonlinear probability $1-\prod_s(1-p_{s,u})$. We also relax $x_{s,i}\in\{0,1\}$ to $x_{s,i}\in[0,1]$ and solve linear programming to obtain fractional $(x^*,y^*)$.

\subsection{Randomized rounding (RR) with feasibility repair}
In our ILP formulation, we solve the relaxed LP and get a fractional allocation $(x^*,y^*)$ where $x^*_{s,i}\in[0,1]$ is the fractional probability of assigning a slot $s$ to the product $i$. As the LP solution is not integral, we use randomized rounding \cite{barahona2005near,raghavan1987randomized} followed by a repair phase to ensure feasibility with respect to budget, disjointness, and balance constraints. First, in the randomized rounding phase, each slot is independently assigned to at most one product. Specifically, we sample a product label $L_s$ according to the fractional values $\{x^*_{s,i}\}$. With probability $x^*_{s,i}$, the slot is assigned to product $i$; with probability $1 - \sum_i x^*_{s,i}$, the slot remains unassigned. Next, in the budget repair phase, some products may exceed their allocated budget (i.e., $|\mathcal{S}_i| > k_i$). To fix this, we iteratively remove excess slots that contribute the least marginal influence from the products until the budget constraint is satisfied. In the balance computation phase, we calculate the estimated influence $\widehat{\mathcal{I}}_i$ for each product based on the current slot assignments. This step allows us to check whether the influence difference constraint, $|\widehat{\mathcal{I}}_i - \widehat{\mathcal{I}}_j| \leq \theta$, is valid for all pairs of products. Next, in the balance repair phase, if the balance constraint is violated, we perform a swap/shift correction in which we identify the product with the maximum influence ($p_{\max}$) and the product with the minimum influence ($p_{\min}$), then attempt to reassign a slot from $p_{\max}$ to $p_{\min}$. The chosen slot is the one that maximizes the net gain in balancing, i.e., minimizes the loss in $p_{\max}$ while maximizing the gain in $p_{\min}$. This process is repeated until either the balance constraint is satisfied or no further beneficial swaps are possible. Finally, an allocation $(\mathcal{S}_1, \mathcal{S}_2, \dots, \mathcal{S}_\ell)$ will be return.

\begin{algorithm}[h!]
\scriptsize
\caption{LP-Relaxation + RR with Balance Repair}
\label{alg:lp-rounding}
\KwIn{LP solution $(x^*,y^*)$, Billboard slots $\mathcal{BS}$, products $\mathcal{P} = \{p_1,p_2,\dots,p_\ell\}$, budgets $\{k_1,\dots,k_\ell\}$, threshold $\theta$, probabilities $p_{s,u}$, sets $\mathcal{U}_i$}
\KwOut{Integral allocation $(\mathcal{S}_1, \mathcal{S}_2,\dots,\mathcal{S}_\ell)$}

\BlankLine
$\mathcal{S}_i \gets \emptyset$ for all $i$; \tcp*{Initialize integral sets}

\BlankLine
\textbf{(A) Per-slot randomized rounding}\;
\For{each slot $s\in \mathcal{BS}$}{
    $\pi_0 \gets \max\{0,\, 1 - \sum_{i\in \mathcal{P}} x^*_{s,i}\}$  \tcp*{probability of leaving $s$ unassigned}
    Sample a label $L_s \in \mathcal{P} \cup \{0\}$ with $\Pr[L_s=i]=x^*_{s,i}$ and $\Pr[L_s=0]=\pi_0$\;
    \If{$L_s \in \mathcal{P}$}{
        $\mathcal{S}_{L_s} \gets \mathcal{S}_{L_s} \cup \{s\}$\;
    }
}

\BlankLine
\textbf{(B) Budget repair}\;
\For{each $i\in \mathcal{P}$}{
    \While{$|\mathcal{S}_i| > k_i$}{
        Choose $s \in \mathcal{S}_i$ with minimum loss: 
        $\text{loss}_i(s) \gets \sum_{u\in \mathcal{U}_i} \min\{p_{s,u},\,1 - \hat{y}_{u,i}\}$, where $\hat{y}_{u,i}$ is current coverage estimate\;
        Remove $s$ from $\mathcal{S}_i$\;
    }
}
\BlankLine
\textbf{(C) Balance computation}\;
\ForEach{$i\in \mathcal{P}$}{
    Estimate influence $\widehat{\mathcal{I}}_i \gets \sum_{u\in \mathcal{U}_i} \min\left\{1,\ \sum_{s\in \mathcal{S}_i} p_{s,u}\right\}$\;
}
\BlankLine
\textbf{(D) Balance repair (swap/shift)}\;
\While{$\exists~ i,j$ with $|\widehat{\mathcal{I}}_i - \widehat{\mathcal{I}}_j| > \theta$}{
    $p_{\max} \gets \arg\max_i \widehat{\mathcal{I}}_i$; \quad $p_{\min} \gets \arg\min_i \widehat{\mathcal{I}}_i$\;
    Find $s^\star \in \mathcal{S}_{p_{\max}}$ maximizing
    \[
    \Delta(s) = \underbrace{\sum_{u\in U_{p_{\min}}} \delta_{p_{\min}}(u;s)}_{\text{gain at }p_{\min}}
    - \underbrace{\sum_{u\in U_{p_{\max}}} \delta_{p_{\max}}(u;s)}_{\text{loss at }p_{\max}},
    \]
    where $\delta_{i}(u;s) = \min\{p_{s,u},\, \max(0,\,1 - \sum_{t\in \mathcal{S}_i\setminus\{s\}} p_{t,u})\}$\;
    \If{$\Delta(s^\star) \le 0$}{\textbf{break}; \tcp*{no beneficial shift, stop}}
     $\mathcal{S}_{p_{\max}} \gets S_{p_{\max}}\setminus\{s^\star\}$, $\mathcal{S}_{p_{\min}} \gets \mathcal{S}_{p_{\min}}\cup\{s^\star\}$\;
    Update $\widehat{\mathcal{I}}_{p_{\max}}, \widehat{\mathcal{I}}_{p_{\min}}$\;
}
\Return{$(\mathcal{S}_1, \mathcal{S}_2,\dots,\mathcal{S}_\ell)$}\;
\end{algorithm}

\paragraph{\textbf{Complexity Analysis.}} Now, we analyze the time and space requirement of Algorithm \ref{alg:lp-rounding}.
First, solving LP relaxation takes polynomial time in the number of slots and products, i.e., $\mathcal{O}(|\mathcal{BS}| \cdot \ell)$ with standard LP solvers. Next, in the randomized rounding phase (line no. $2$ to $7$), it assigns each slot independently and therefore runs in $\mathcal{O}(|\mathcal{BS}|\cdot \ell)$. In the budget repair phase (line no. $9$ to $12$), we may at most remove all excess slots once, giving $\mathcal{O}(|\mathcal{BS}|\cdot \ell)$ time in the worst case. In line $13$ to $15$, computing the product-specific influences takes $\mathcal{O}(|\mathcal{BS}|\cdot \bar{d} \cdot \ell)$, where $\bar{d}$ is the number of tuples in the trajectory database. In the balance repair phase (line no. $16$ to $23$), performs iterative slot shifts; in the worst case, each slot may be moved once, giving another $\mathcal{O}(|\mathcal{BS}|\cdot \bar{d} \cdot \ell)$. Hence, the total running time is bounded by $\mathcal{O}( \text{LP-solve} + |\mathcal{BS}|\cdot \ell \cdot \bar{d})$.

\par The Algorithm \ref{alg:lp-rounding} requires storage for the fractional LP solution $(x^*, y^*)$, which has $\mathcal{O}(|\mathcal{BS}|\cdot \ell)$ variables. Additionally, we maintain the slot allocation sets $(\mathcal{S}_1, \mathcal{S}_2, \dots, \mathcal{S}_\ell)$, requiring $\mathcal{O}(|\mathcal{BS}|)$ space. Storing user-specific influence values requires at most $\mathcal{O}(|\mathcal{U}|\cdot \ell)$. Therefore, the total space complexity is $\mathcal{O}\big(|\mathcal{BS}|\cdot \ell + |\mathcal{U}|\cdot \ell \big)$ or equivalently $\mathcal{O}(|\mathcal{BS}|\cdot \ell)$. Hence, Theorem \ref{Th:CA2} holds.

\begin{theorem}\label{Th:CA2}
The time and space requirements for Algorithm \ref{alg:lp-rounding} will be $\mathcal{O}( \text{LP-solve} + |\mathcal{BS}|\cdot \ell \cdot \bar{d})$ and $\mathcal{O}(|\mathcal{BS}|\cdot \ell)$, respectively.
\end{theorem}

\subsection{Greedy Heuristic with Balance Correction}
The Algorithm \ref{alg:greedy-balance} starts by initializing empty slot sets $\mathcal{S}_i$ for each product $i \in \mathcal{P}$. First, in line numbers $2$ to $15$, in the greedy allocation phase for each product, we sample a random subset of $\left\lceil \frac{|\mathcal{BS}|}{k} \cdot \log \tfrac{1}{\epsilon} \right\rceil$ many candidate slots from $\mathcal{BS}$. The candidate slots assign the slot to the products with the highest influence gain. This process continues until no product has remaining budget. Next in line $16$ to $30$, first the influence values $\mathcal{I}_i(\mathcal{S}_i)$ are calculated for all products. If the influence difference between any two products exceeds the threshold $\theta$, $p_{max}$ and $p_{min}$ are identified. Next, we evaluate each slot $s \in p_{max}$ and calculate its potential benefit if reassigned to $p_{min}$. The best slot $s^* \in p_{max}$ that maximizes balance improvement is selected and re-assigned from $p_{max}$ to $p_{min}$. This correction process is repeated until the influence differences among all products are within $\theta$. Finally, the algorithm outputs an allocation $(\mathcal{S}_1, \mathcal{S}_2, \dots, \mathcal{S}_\ell)$. This method guarantees feasibility by satisfying constraints, while achieving a near-optimal aggregated influence through greedy selection and local adjustments.

\begin{algorithm}[h!]
\caption{Greedy Heuristic with Balance Correction}
\label{alg:greedy-balance}
\scriptsize
\KwData{Billboard slots $\mathcal{BS}$, products $\mathcal{P} = \{p_1,p_2,\dots,p_\ell\}$, budgets $\{k_1,\dots,k_\ell\}$, threshold $\theta$, error parameter $\epsilon$}
\KwResult{Allocation $(\mathcal{S}_1, \dots, \mathcal{S}_\ell)$ of slots to products}
Initialize $\mathcal{S}_i \gets \emptyset$ for all $i \in \mathcal{P}$, and $S^{'} \gets \emptyset$ (set of used slots)\;

\BlankLine
\textbf{--- Greedy Allocation Phase ---}\;
\For{each product $i \in \mathcal{P}$}{
    \While{$|\mathcal{S}_i| < k_i$}{
        Set $k \gets \max\{1, \lceil |\mathcal{BS}| \cdot 0.1 \rceil\}$\;
        Sample size: $r \gets \left\lceil \frac{|\mathcal{BS}|}{k} \cdot \log \tfrac{1}{\epsilon} \right\rceil$\;
        Select candidate subset $\mathcal{R} \subseteq \mathcal{BS}$ uniformly at random, with $|\mathcal{R}| = \min(r, |\mathcal{BS}|)$\;
        \For{slot $s \in \mathcal{R}$}{
            \If{$s \notin S^{'}$ and $s \notin \mathcal{S}_i$ and $\text{cost}(s) \leq k_i$}{
                Compute marginal gain: $\Delta \mathcal{I}_i(s) = \mathcal{I}_i(\mathcal{S}_i \cup \{s\}) - \mathcal{I}_i(\mathcal{S}_i)$\;
            }
        }
        Choose $s^* = \arg\max_{s \in \mathcal{R}} \Delta \mathcal{I}_i(s)$\;
        \If{$s^* \neq \emptyset$}{
            Assign slot: $\mathcal{S}_i \gets \mathcal{S}_i \cup \{s^*\}$, $S^{'} \gets S^{'} \cup \{s^*\}$\;
        }
        \Else{break}
    }
}

\BlankLine
\textbf{--- Balance Correction Phase ---}\;
Compute $\mathcal{I}_i(\mathcal{S}_i)$ for all $i \in \mathcal{P}$\;
\While{$\max_i \mathcal{I}_i(\mathcal{S}_i) - \min_j \mathcal{I}_j(\mathcal{S}_j) > \theta$}{
    $p_{max} \gets \arg\max_i \mathcal{I}_i(\mathcal{S}_i)$, \quad $p_{min} \gets \arg\min_i \mathcal{I}_i(\mathcal{S}_i)$\;
    \For{$s \in \mathcal{S}_{p_{max}}$}{
        Compute gain if reassigned: \\
        $gain_{min} = \mathcal{I}_{p_{min}}(\mathcal{S}_{p_{min}} \cup \{s\}) - \mathcal{I}_{p_{min}}(\mathcal{S}_{p_{min}})$\;
        $loss_{max} = \mathcal{I}_{p_{max}}(\mathcal{S}_{p_{max}}) - \mathcal{I}_{p_{max}}(\mathcal{S}_{p_{max}} \setminus \{s\})$\;
        $\Delta(s) = gain_{min} - loss_{max}$\;
    }
    Select $s^* = \arg\max_s \Delta(s)$\;
    \If{$s^* \neq \emptyset$}{
        Reassign: $\mathcal{S}_{p_{max}} \gets \mathcal{S}_{p_{max}} \setminus \{s^*\}$, \quad $\mathcal{S}_{p_{min}} \gets \mathcal{S}_{p_{min}} \cup \{s^*\}$\;
        Update influences $\mathcal{I}_i(\mathcal{S}_i)$ for all $i$\;
    }
    \Else{break}
}

\Return{$(\mathcal{S}_1, \dots, \mathcal{S}_\ell)$}\;
\end{algorithm}


\paragraph{\textbf{Complexity Analysis.}}
Now, we analyze the time and space requirements of Algorithm \ref{alg:greedy-balance}.
In Line No. $1$, initializing the allocation set will take $\mathcal{O}(\ell)$ time. In Line No. $3$ \texttt{for loop} will execute for $\mathcal{O}(\ell)$ and the \texttt{while loop} will execute till the demand of each product is satisfied. So, in Line No. $3$ to $15$ in each iteration, computing the marginal gain for every feasible pair of product $i \in \mathcal{P}$ and available slot $s \in \mathcal{BS}$ will take $\mathcal{O}(\left\lceil \frac{|\mathcal{BS}|}{k} \cdot \log \tfrac{1}{\epsilon} \right\rceil \cdot \ell \cdot \bar{d})$ time per iteration. Since at most $\sum_{i=1}^{\ell} k_i$ slots are assigned, the total time requirements will be of $\mathcal{O}(\left\lceil \frac{|\mathcal{BS}|}{k} \cdot \log \tfrac{1}{\epsilon} \right\rceil \cdot \ell \cdot \bar{d} \cdot \sum_{i=1}^{\ell} k_i)$. Next, in Line No. $17$ to $30$ for each reassignment step requiring at most $\mathcal{O}(k_{\max})$ evaluations, where $k_{\max} = \max_i k_i$. Since this process may repeat for up to $\mathcal{O}(|\mathcal{BS}|)$ iterations, the overall cost of this phase is $\mathcal{O}(|\mathcal{BS}|\cdot k_{\max})$. So, combining both phases, the overall time complexity is $\mathcal{O}(\left\lceil \frac{|\mathcal{BS}|}{k} \cdot \log \tfrac{1}{\epsilon} \right\rceil \cdot \ell \cdot \bar{d} \cdot \sum_{i=1}^{\ell} k_i + |\mathcal{BS}|\cdot k_{\max})$. The additional space requirements for allocation set $(\mathcal{S}_1,\dots,\mathcal{S}_\ell)$ will be $\mathcal{O}(|\mathcal{BS}|)$. The influence values $\mathcal{I}_i(\mathcal{S}_i)$ must be stored for each product takes $\mathcal{O}(\ell)$ space. So, the additional space requirement will be $\mathcal{O}(|\mathcal{BS}| + \ell)$. Hence, Theorem \ref{Th:CA1} holds.
\begin{theorem}\label{Th:CA1}
The time and space requirements for Algorithm \ref{alg:greedy-balance} are  $\mathcal{O}(\left\lceil \frac{|\mathcal{BS}|}{k} \cdot \log \tfrac{1}{\epsilon} \right\rceil \cdot \ell \cdot \bar{d} \cdot \sum_{i=1}^{\ell} k_i + |\mathcal{BS}|\cdot k_{\max})$ and $\mathcal{O}(|\mathcal{BS}| + \ell)$, respectively.
\end{theorem}
\section{Experimental Setup} \label{Sec:Experiments}
This section describes the experimental setup for the experiments. Initially, we start by reporting the datasets used in our experiments.
\paragraph{\textbf{Dataset Description}.}
We used two real-world datasets from New York City (NYC)\footnote{\url{https://www.nyc.gov/site/tlc/about/tlc-trip-record-data.page}} and Los Angeles (LA)\footnote{\url{https://github.com/Ibtihal-Alablani}}, previously used in related studies \cite{ali2022influential,ali2023influential,ali2024influential,ali2024regret}. The NYC dataset contains 227,428 check-ins (April 2012 to February 2013), and the LA dataset has 74,170 check-ins across 15 streets. Billboard data were obtained from LAMAR\footnote{\url{http://www..lamar.com/InventoryBrowser}}, yielding 1,031,040 slots for NYC and 2,135,520 slots for LA.
\paragraph{\textbf{Key Parameters.}}
All the parameters are summarized in Table \ref{Key-parameters}. First, the Demand-Supply Ratio $(\alpha)$ defines the ratio $\alpha = \sigma^{\mathcal{P}} / \sigma^{*}$ as the proportion of global demand $\big(\sigma^{\mathcal{P}} = \sum_{i=1}^{n} \sigma_i\big)$ to total supply $(\sigma^{*} = |\mathcal{BS}|)$. We evaluate $\alpha$ at four levels: $40\%, 60\%, 80\%, 100\%$. Second, the Average-Individual Demand Ratio $(\beta)$ defines $\beta = \sigma^{\mathcal{P}''} / \sigma^{*}$ as the ratio of average individual demand to total supply, where $\sigma^{\mathcal{P}''} = \sigma^{\mathcal{P}} / |\mathcal{P}|$ denotes the average demand per product. The parameter $\beta$ is used to scale the demand of individual products. Third, the demand for each product is computed as $\sigma = \lfloor \omega \cdot \sigma^{*} \cdot \beta \rfloor$, where $\omega \in [0.8,1.2]$ is a random factor used to capture variations in product payments. Fourth, the threshold parameter $(\theta)$ controls fairness tolerance. It defines the maximum allowed difference between the highest and lowest product influence. Fifth, $\epsilon$ controls the probability of error in random sampling. We vary one parameter in each experiment and set the remaining parameters in their default settings (highlighted in bold). To solve the LP, we use the HiGHS\footnote{\url{https://highs.dev/}} solver. All the Python codes are executed in HP Z4 workstations with 64 GB RAM and an Xeon(R) 3.50 GHz processor.  
\begin{table}[h!]
\caption{\label{Key-parameters} Key Parameters}
\vspace{-0.15 in}
\begin{center}
    \begin{tabular}{ | p{2cm}| p{5.5cm}|}
    \hline
    Parameter & Values  \\ \hline
    $\alpha$ & $40\%, 60\%, 80\%, \textbf{100\%}$   \\ \hline
    $\beta$ & $1\%, 2\%, \textbf{5\%}, 10\%$   \\ \hline
    $\epsilon$ & $0.01, 0.05, \textbf{0.1}, 0.2$   \\ \hline
    $\theta$ & $0.02, \textbf{0.05}, 0.1, 0.2$   \\ \hline
    $\lambda$ & $50m,\textbf{100m},125m,150m$  \\ \hline
    \end{tabular}
\end{center}
\end{table}

\paragraph{\textbf{Baseline Methods.}}
\paragraph{\textit{Random + Balance Correction.}}
In this approach, the billboard slots are selected uniformly at random to allocate to products until the budget constraint is satisfied. Next, the balance correction phase is applied, which ensures that the influence difference constraint is satisfied.
\paragraph{\textit{Top-$k$ + Balance Correction.}}
In this approach, the billboard slots are sorted according to their influence value. Then, sorted slots are assigned to the products individually until the budget constraint is satisfied. Finally, the balance correction is used to satisfy the influence difference.
\paragraph{\textbf{Goals of our Experiments.}} \label{Sec:Research_Questions}
In this study, we want to address the following Research Questions (RQ).
\begin{itemize}
\item \textbf{RQ1}: Varying $\alpha$, $\beta$, how does the influence and runtime change?
\item \textbf{RQ2}: Varying $\theta$, how do the fairness gap and run time changes?
\item \textbf{RQ3}: Varying $\epsilon$ and $\lambda$, how does the influence and run time change?
\end{itemize}
\paragraph{\textbf{Experimental Results and Discussion}}
\paragraph{\textit{Budget, $\epsilon$ Vs. Influence.}}
Figure \ref{Fig:NYC}(a,b,c,d) and Figure \ref{Fig:LA}(a,b,c,d) show the effectiveness of the proposed solution approaches. We have four main observations. First, with increasing demand-supply ratio $(\alpha)$ from $40\%$ to $100\%$ with a fixed value of $\beta$, $\epsilon$, and $\theta$, the influence of all proposed and baseline methods increases. Second, the proposed `LP+RR' approach achieves more influence than the `Greedy Heuristic'. The `LP+RR' approach gives better quality results (higher influence, tighter fairness balance) because it leverages LP relaxation, which approximates the global optimum before rounding. Third, among the baselines, the `Top-$k$' approach performs better than `Random' in terms of influence maximization. Fourth, the`Greedy Heuristic' is faster and scalable, however, at the cost of lower influence compared to `LP+RR'. Figures \ref{Fig:NYC}(g) and \ref{Fig:LA}(g) show that with increasing the error probability parameter $(\epsilon)$ value from $0.01$ to $0.2$, the quality of the solution degrades. The lower $\epsilon$ improves the quality of the solutions; however, the computational cost increases.
\begin{figure*}[!ht]
\centering
\begin{tabular}{cccc}
\includegraphics[scale=0.12]{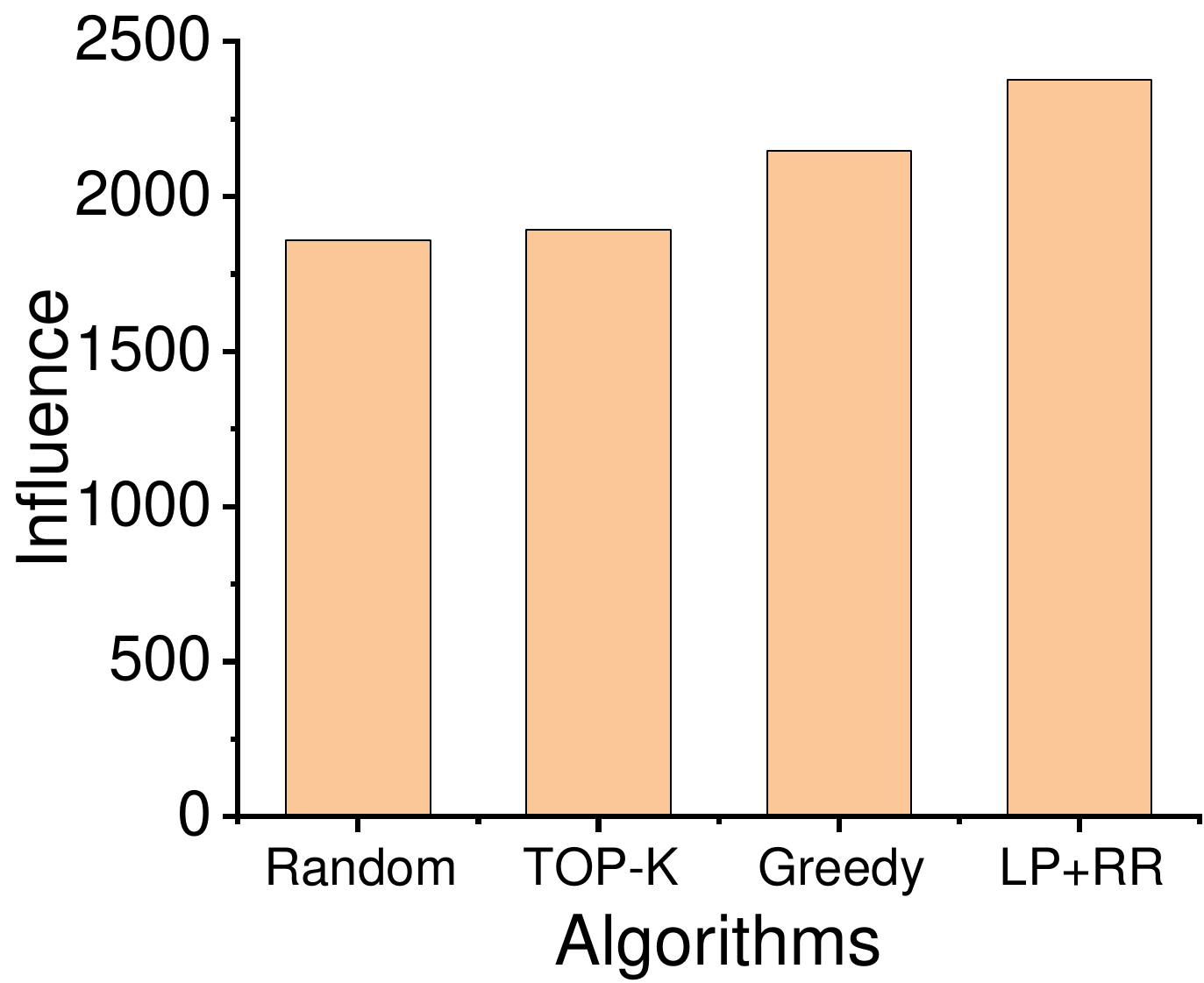} & \includegraphics[scale=0.12]{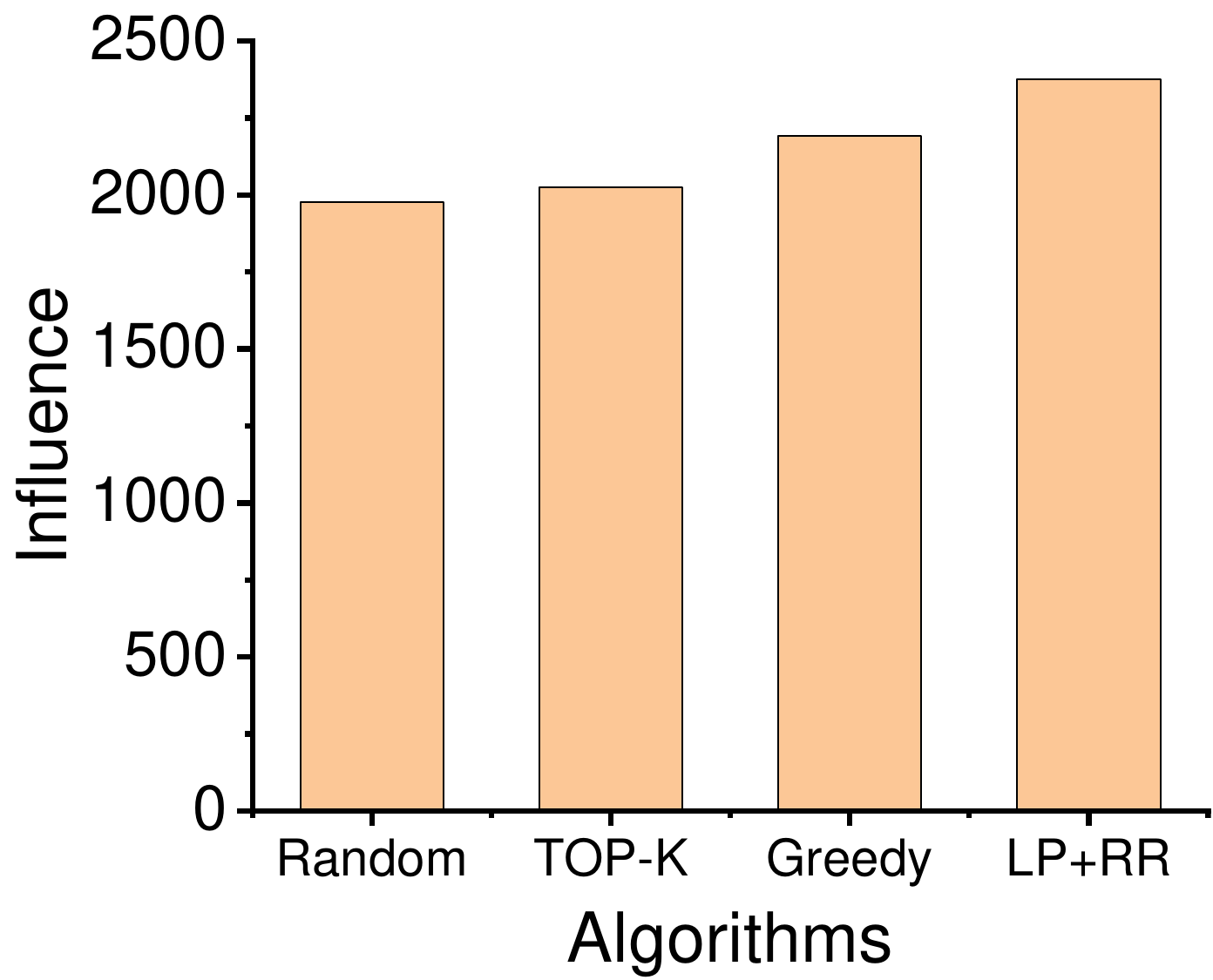} & \includegraphics[scale=0.12]{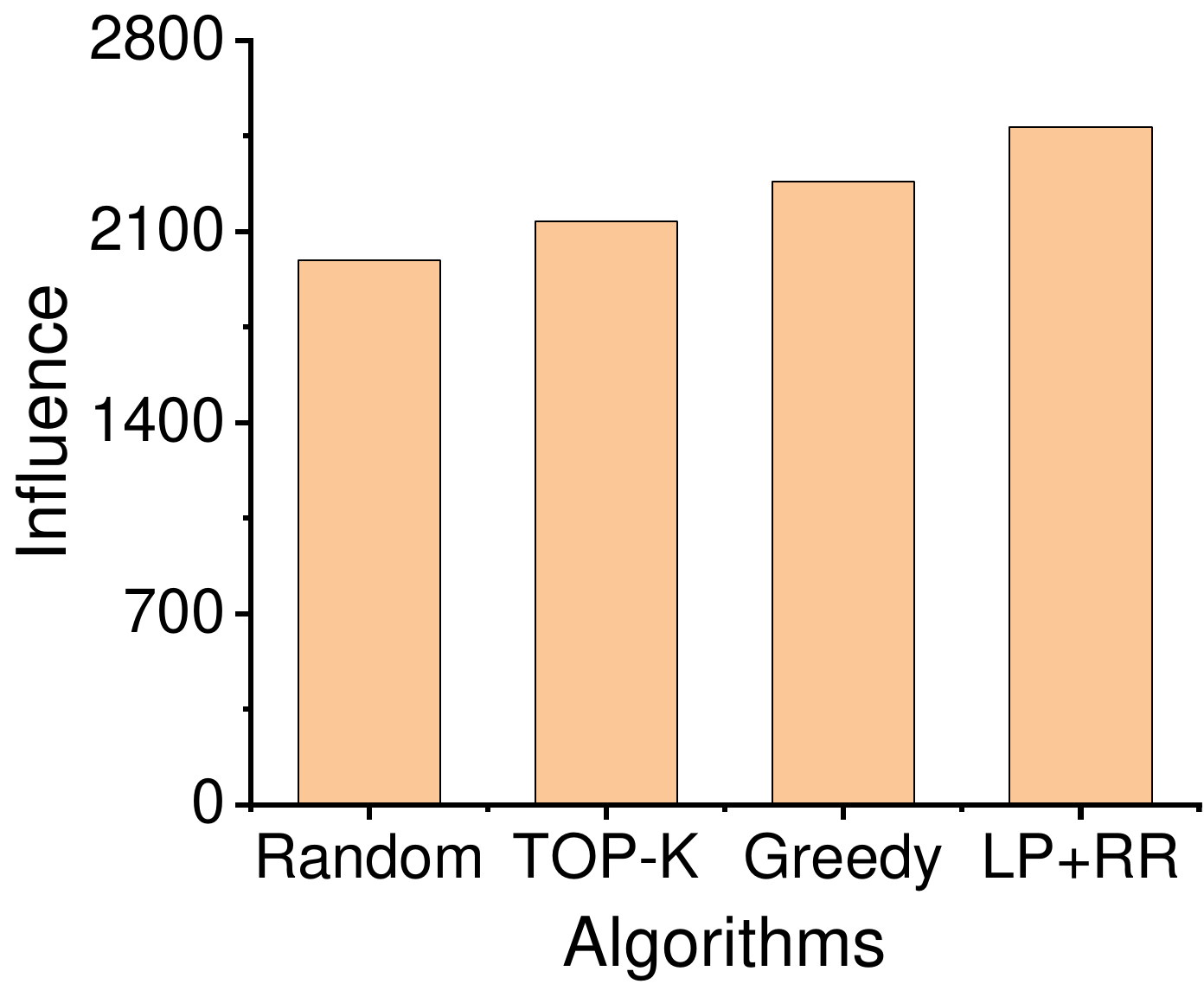} & \includegraphics[scale=0.12]{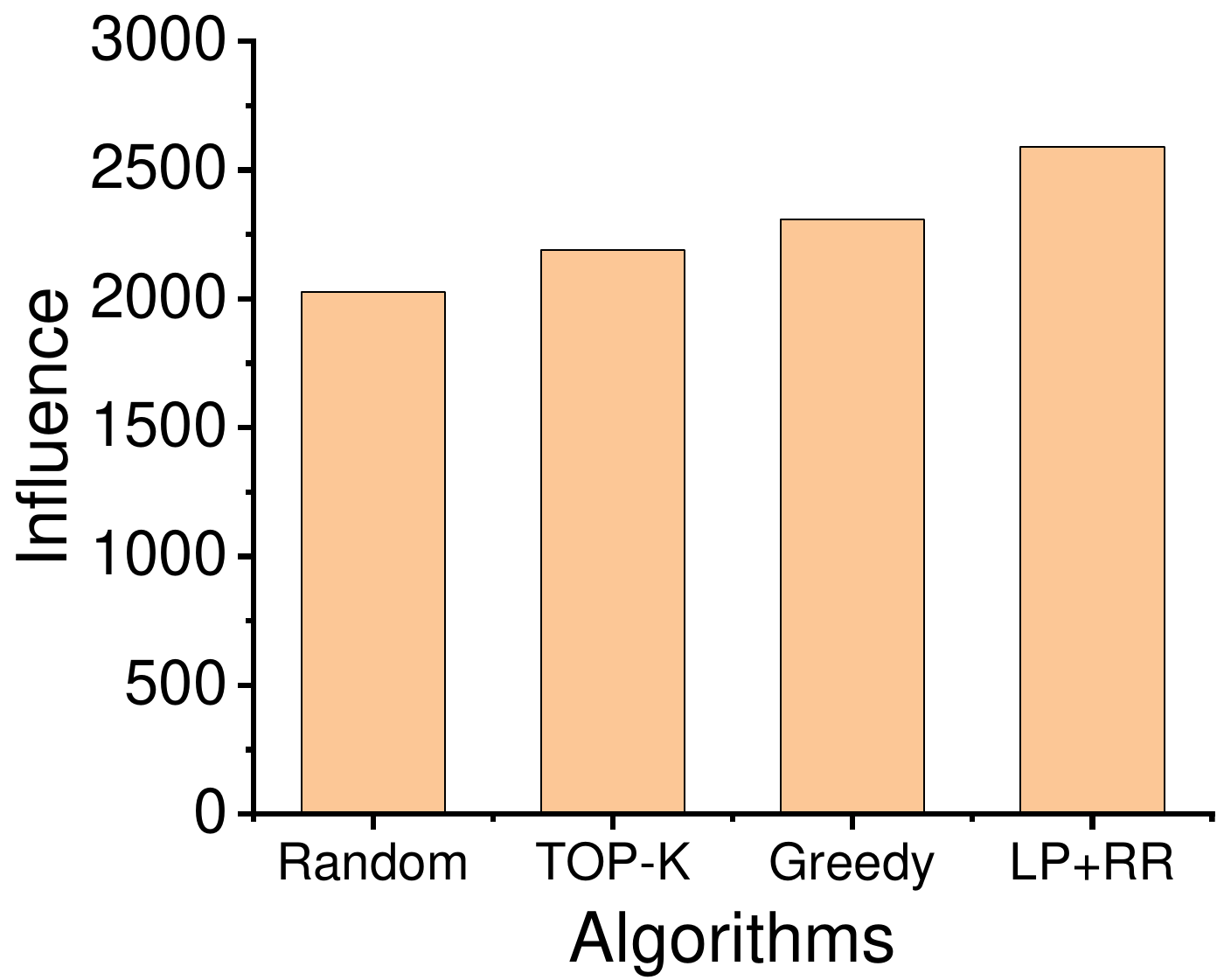} \\
\tiny{(a) $\alpha = 40\%$} &  \tiny{(b) $\alpha = 60\%$} & \tiny{(c) $\alpha = 80\%$} & \tiny{(d)$\alpha = 100\%$}  \\
\includegraphics[scale=0.12]{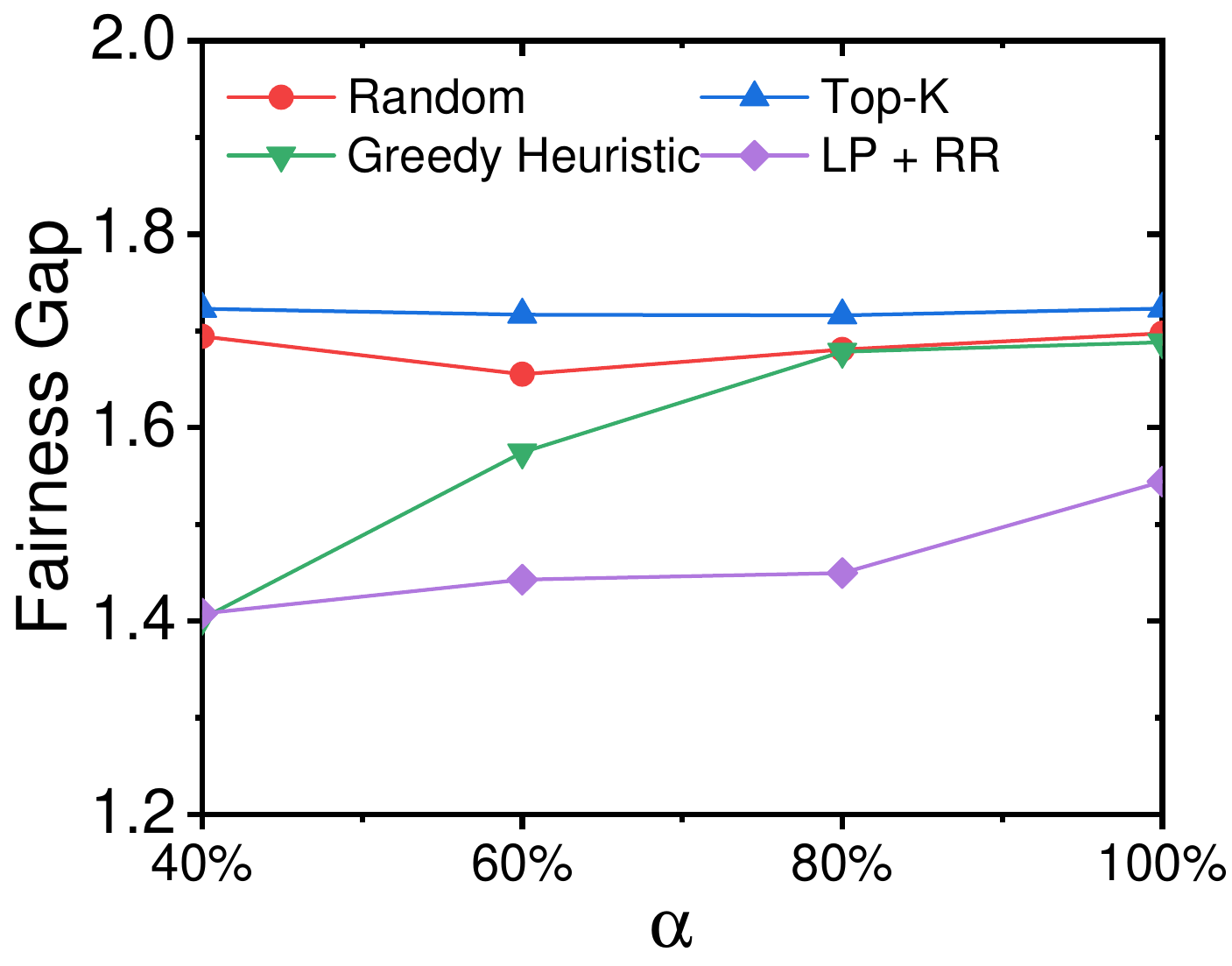} & \includegraphics[scale=0.12]{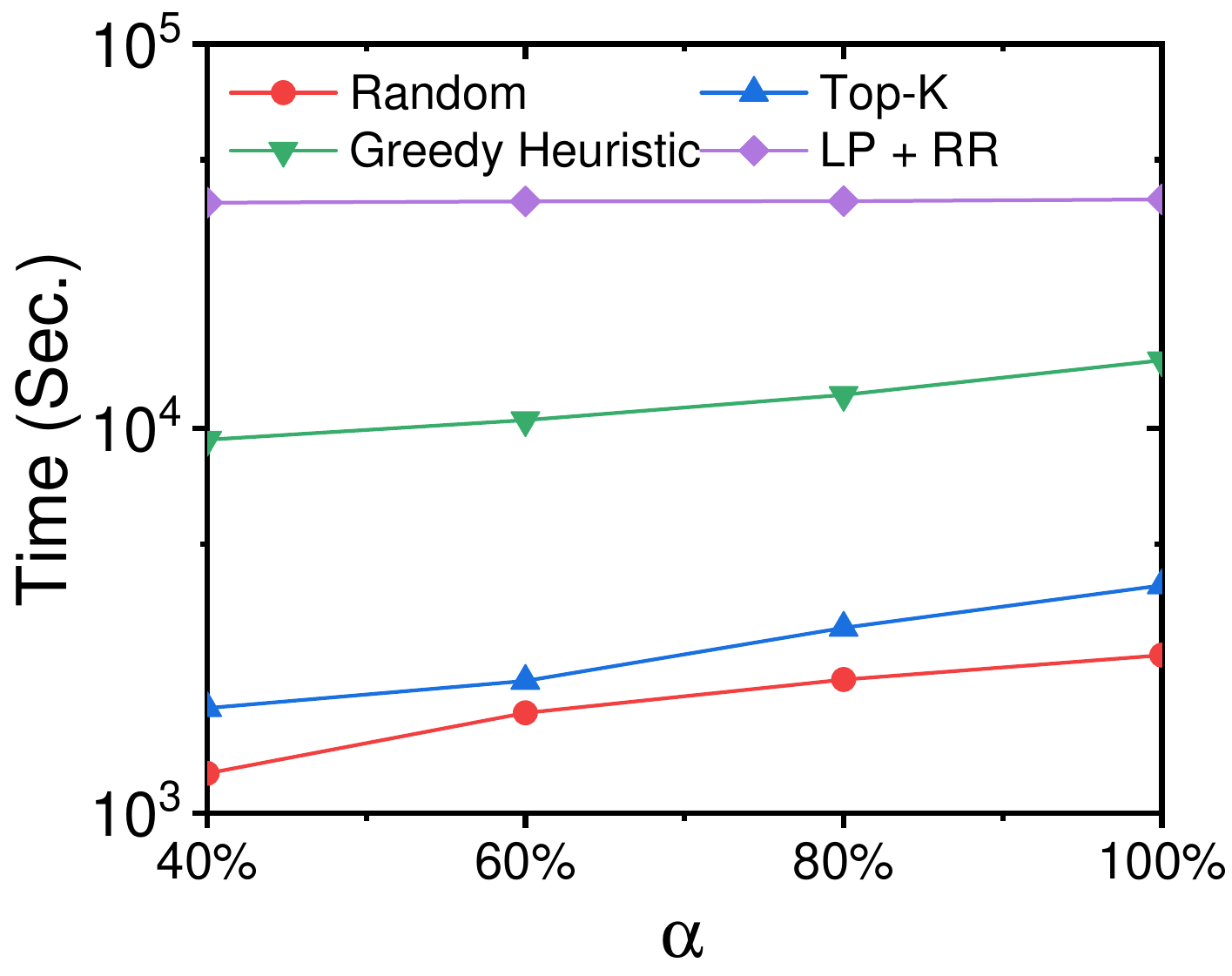} & \includegraphics[scale=0.12]{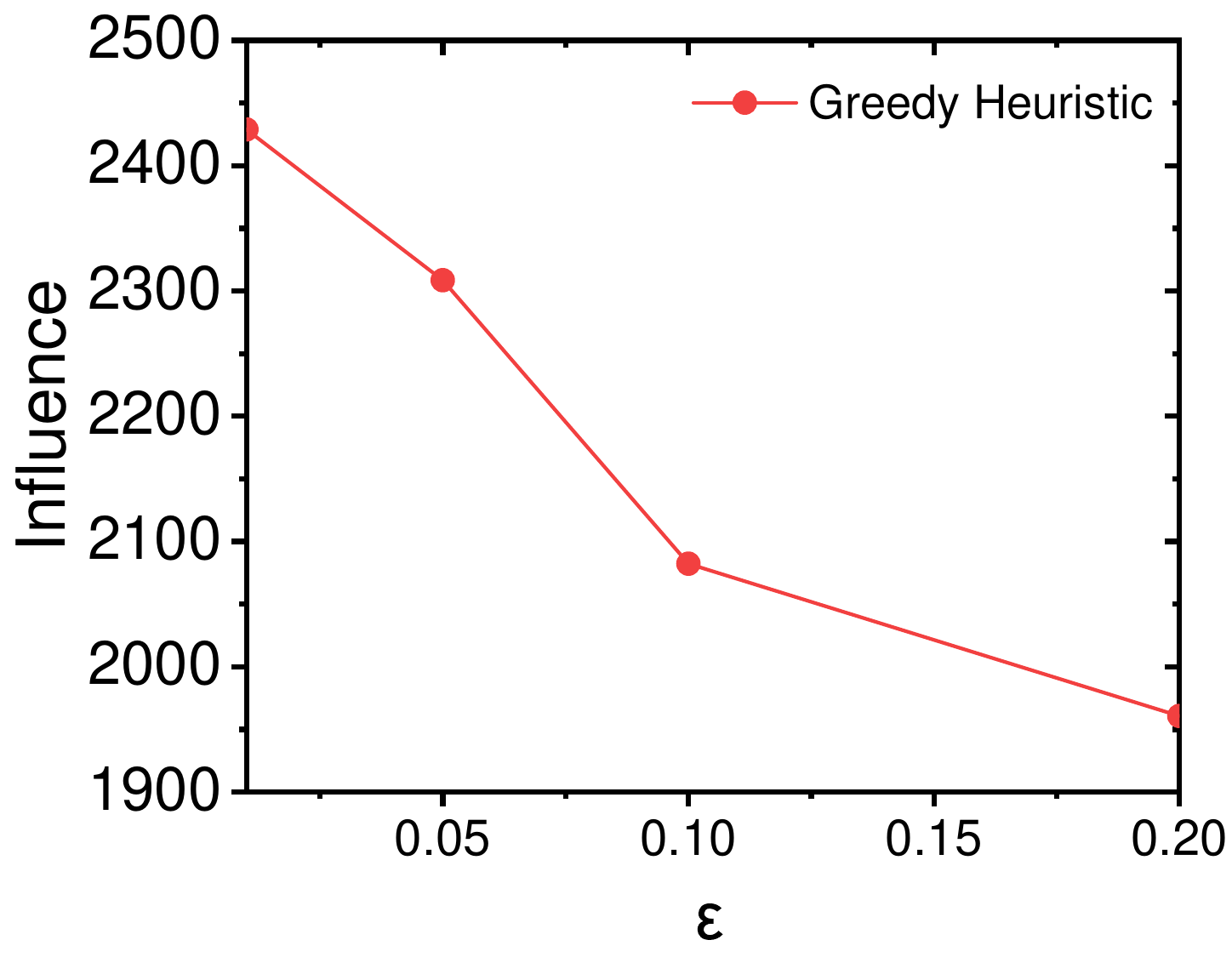} & \includegraphics[scale=0.12]{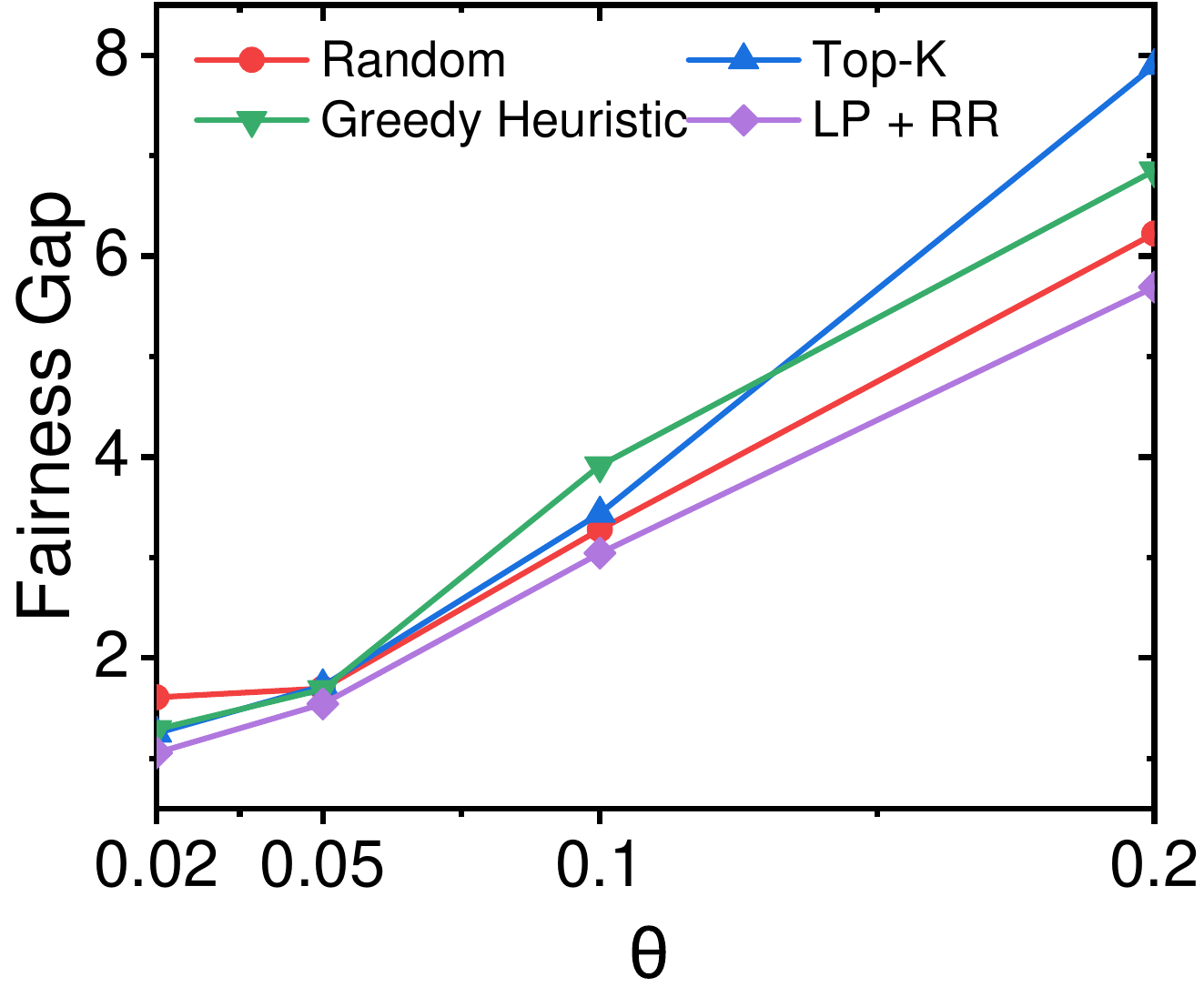} \\
\tiny{(e) $\epsilon =0.1$, $\theta = 0.05$} &  \tiny{(f) $\epsilon =0.1$, $\theta = 0.05$} & \tiny{(g) $\alpha = 100\%$, $\theta = 0.05$} & \tiny{(h) $\alpha = 100\%$, $\beta = 5\%$}  \\
\end{tabular}
\caption{ Algorithms Vs. Influence $(a,b,c,d)$, Varying $\alpha$ Vs. Fairness Gap (e), Time (f), Varying $\epsilon$ Vs. Influence (g), Varying $\theta$ Vs. Fairness Gap (h) with $|\mathcal{P}| = 20$, $\beta = 5\%$ for NYC dataset}
\label{Fig:NYC}
\vspace{-0.25 in}
\end{figure*}

\paragraph{\textit{$\theta$ Vs. Fairness Gap.}}
Figures \ref{Fig:NYC} (h) and \ref{Fig:LA} (h) show the effectiveness of the fairness tolerance parameter $\theta$ in maximizing influence across trajectories. The smaller $\theta$ shows that the difference in influence between the products is very close (strict balance), and the larger $\theta$ allows for larger gaps in influence between the products. With an increase in $\theta$ from $0.02$ to $0.2$, the influence gap for all proposed and baseline methods increases. Figures \ref{Fig:NYC} (e) and \ref{Fig:LA} (e) show that with increasing $\alpha$, the fairness gap also increases due to increasing the budget for the products. The `Top-$k$' achieves the highest influence gap; however, `LP + RR' achieves the lowest influence gap. If the trajectories show a dense overlap, some products will naturally have more influence than others, so we can allow more imbalance $(\text{bigger}~ \theta)$. For example, we say that the imbalance between products should not exceed, say, $ 0.1$ or $10\%$ of the typical coverage per slot.

\begin{figure*}[!ht]
\centering
\begin{tabular}{cccc}
\includegraphics[scale=0.12]{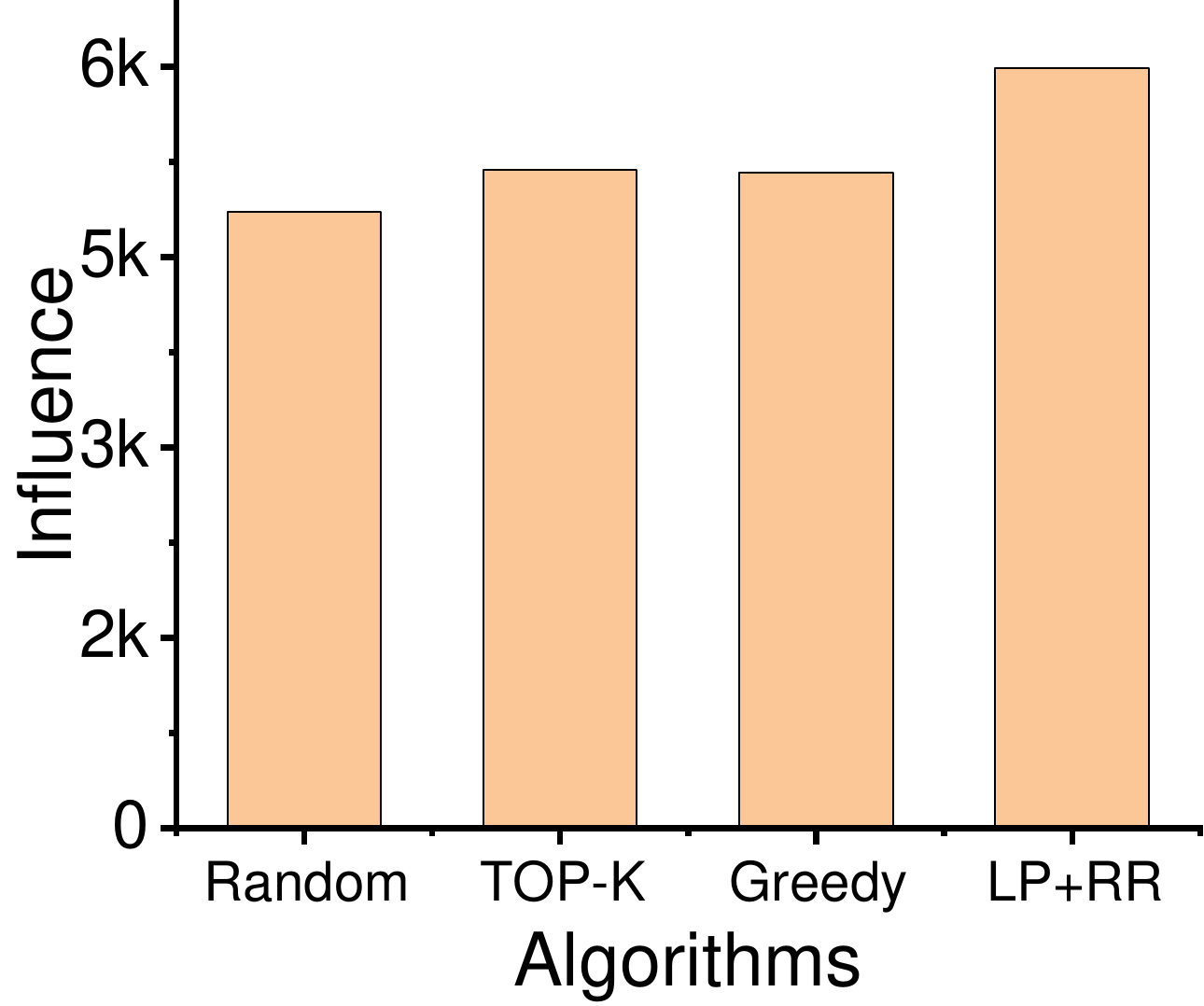} & \includegraphics[scale=0.12]{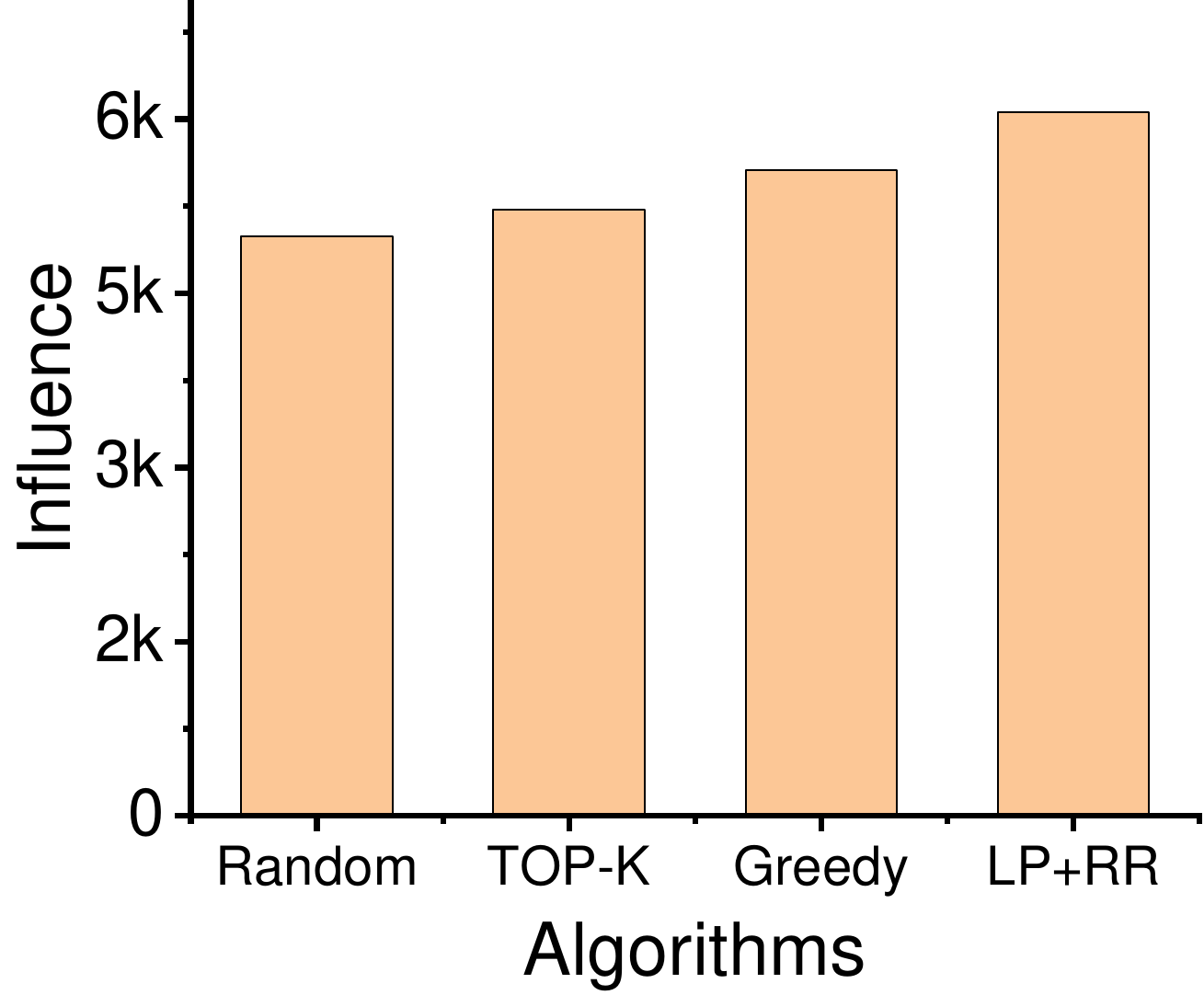} & \includegraphics[scale=0.12]{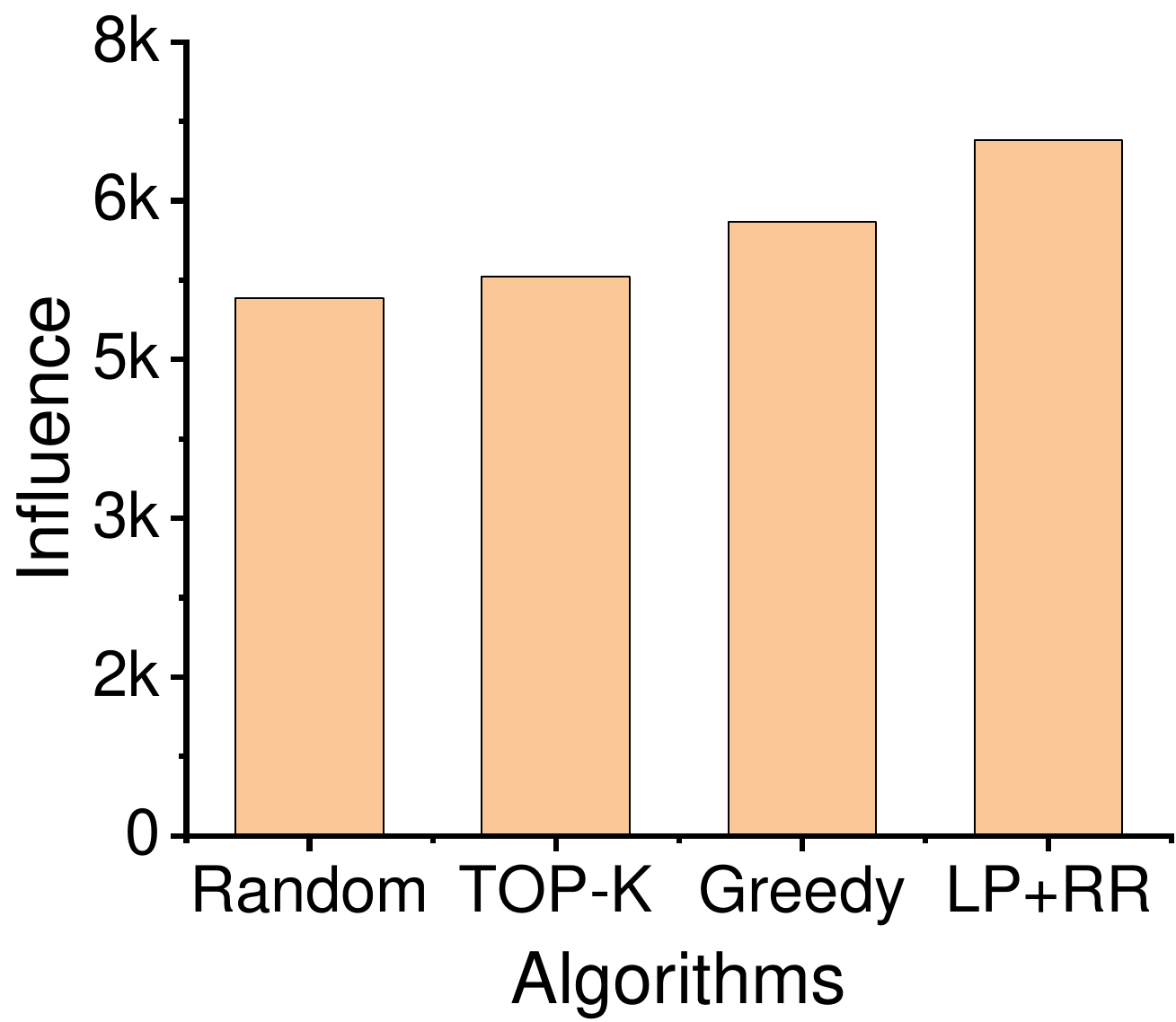} & \includegraphics[scale=0.12]{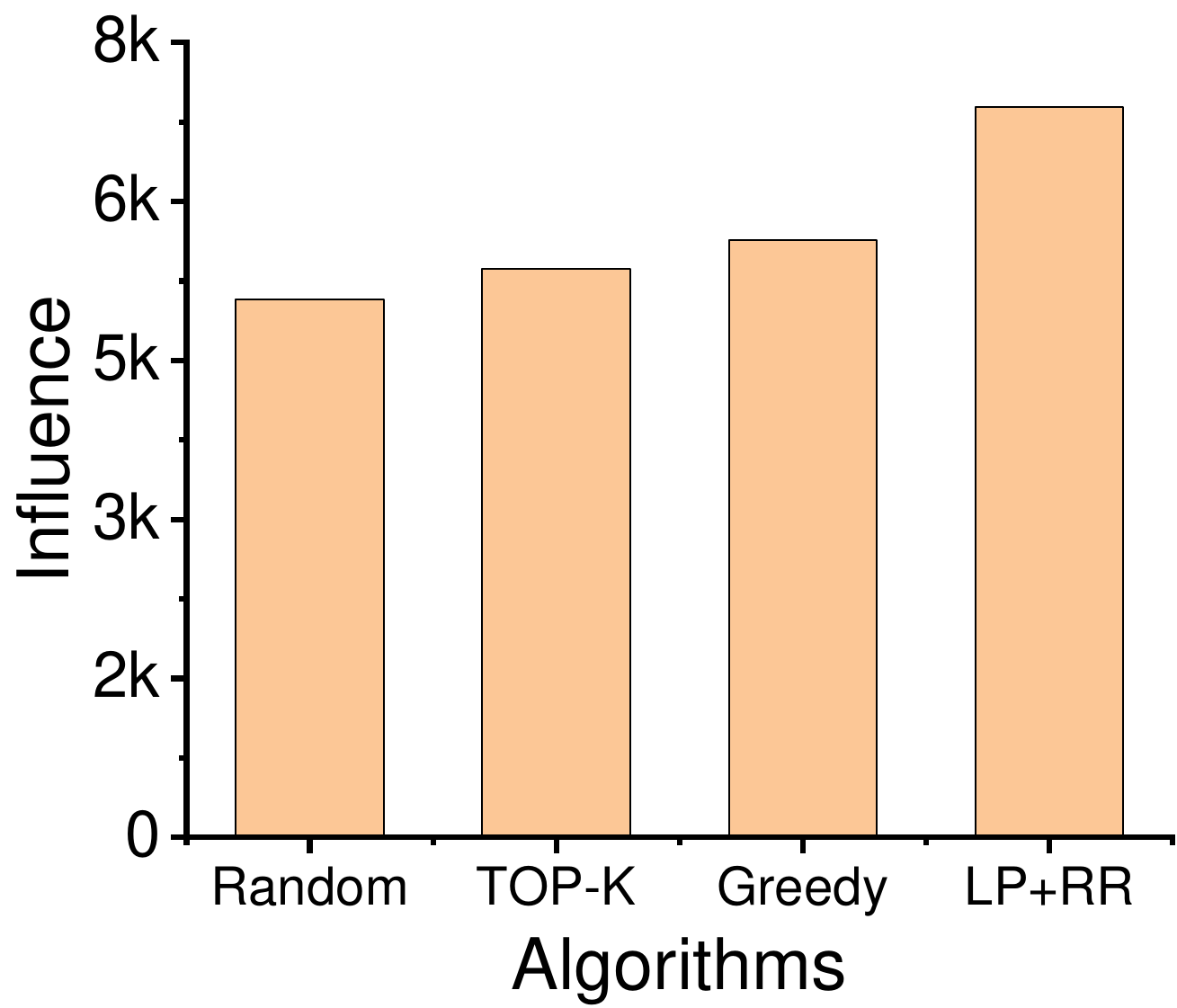} \\
\tiny{(a) $\alpha = 40\%$} &  \tiny{(b) $\alpha = 60\%$} & \tiny{(c) $\alpha = 80\%$} & \tiny{(d)$\alpha = 100\%$}  \\
\includegraphics[scale=0.12]{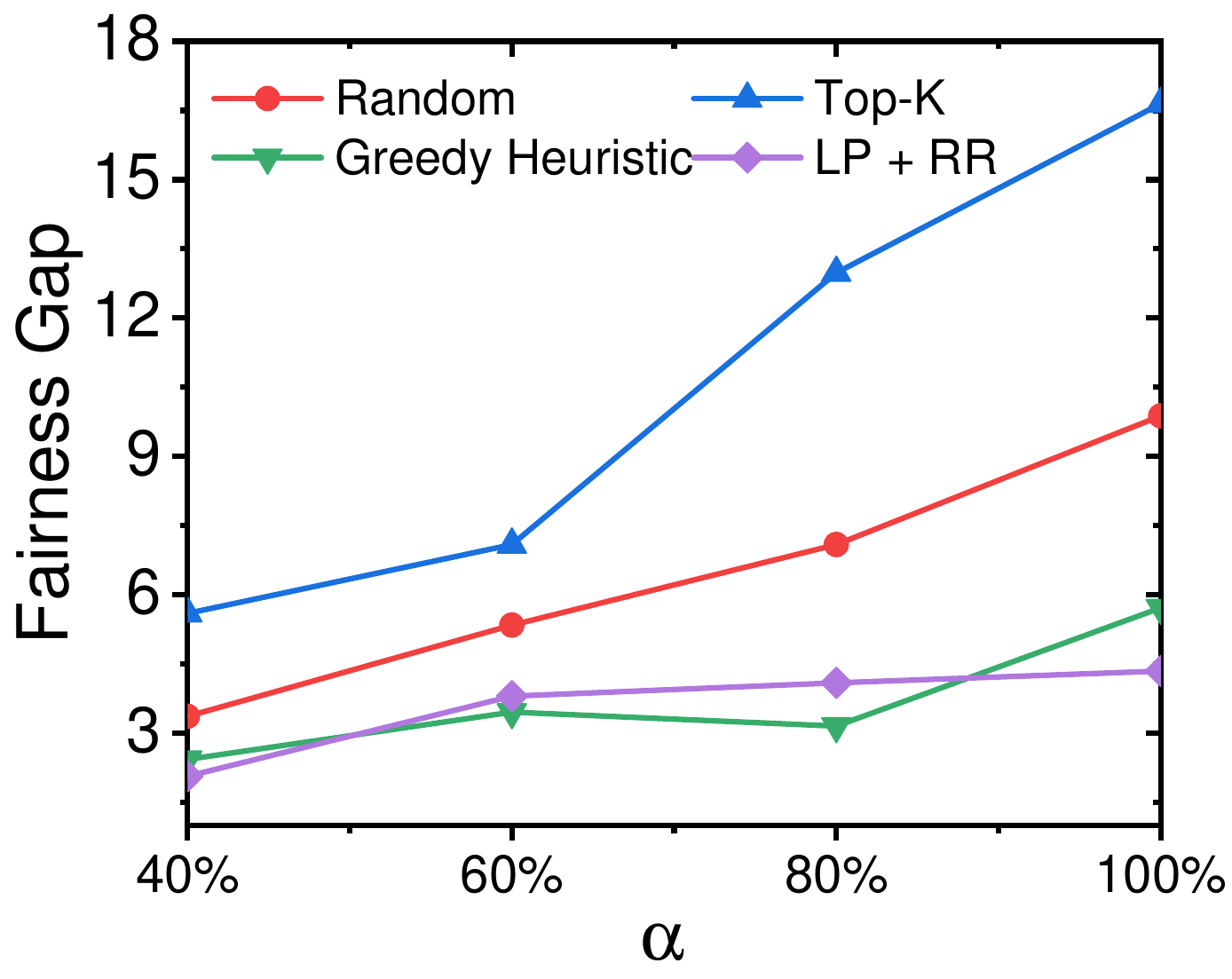} & \includegraphics[scale=0.12]{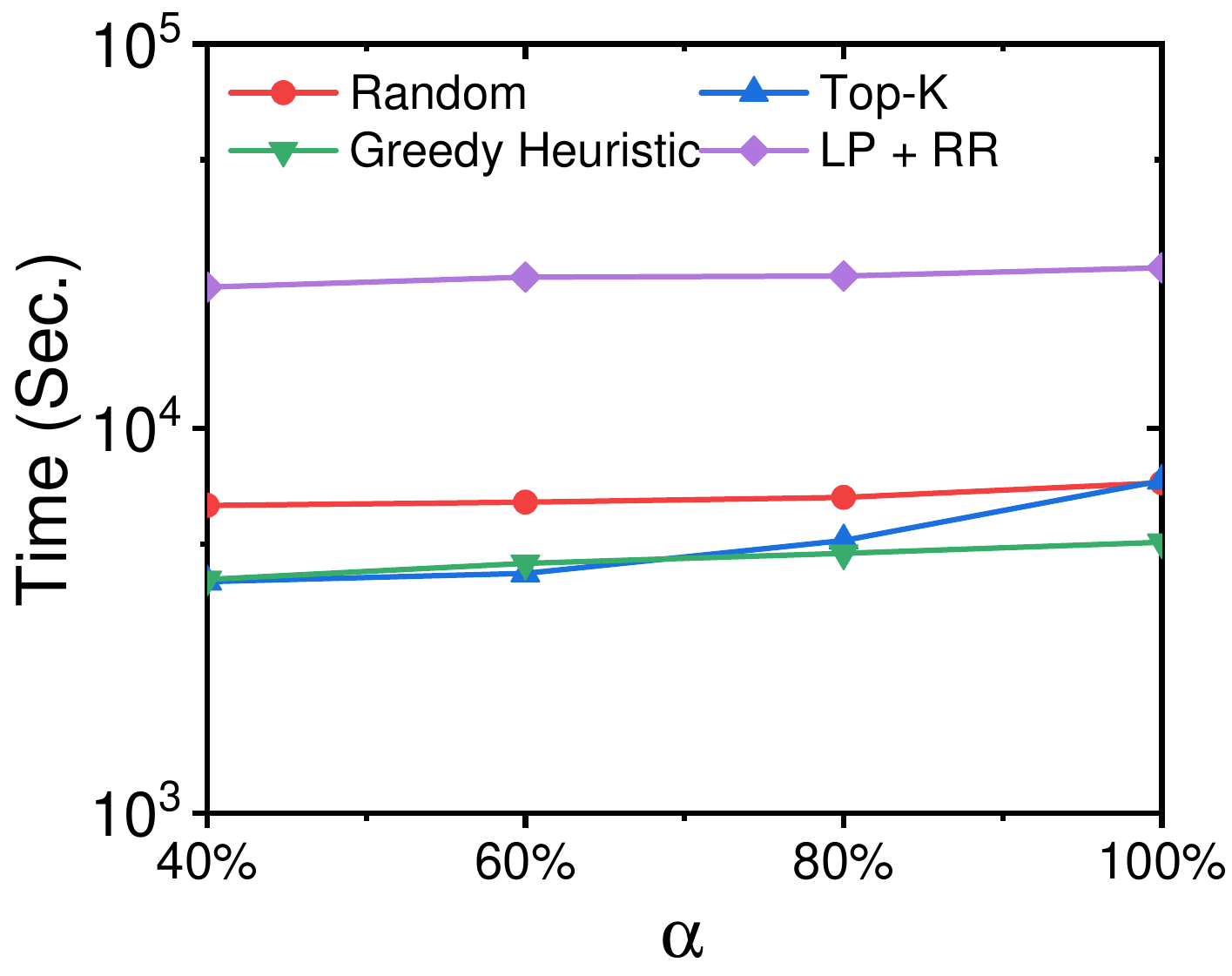} & \includegraphics[scale=0.12]{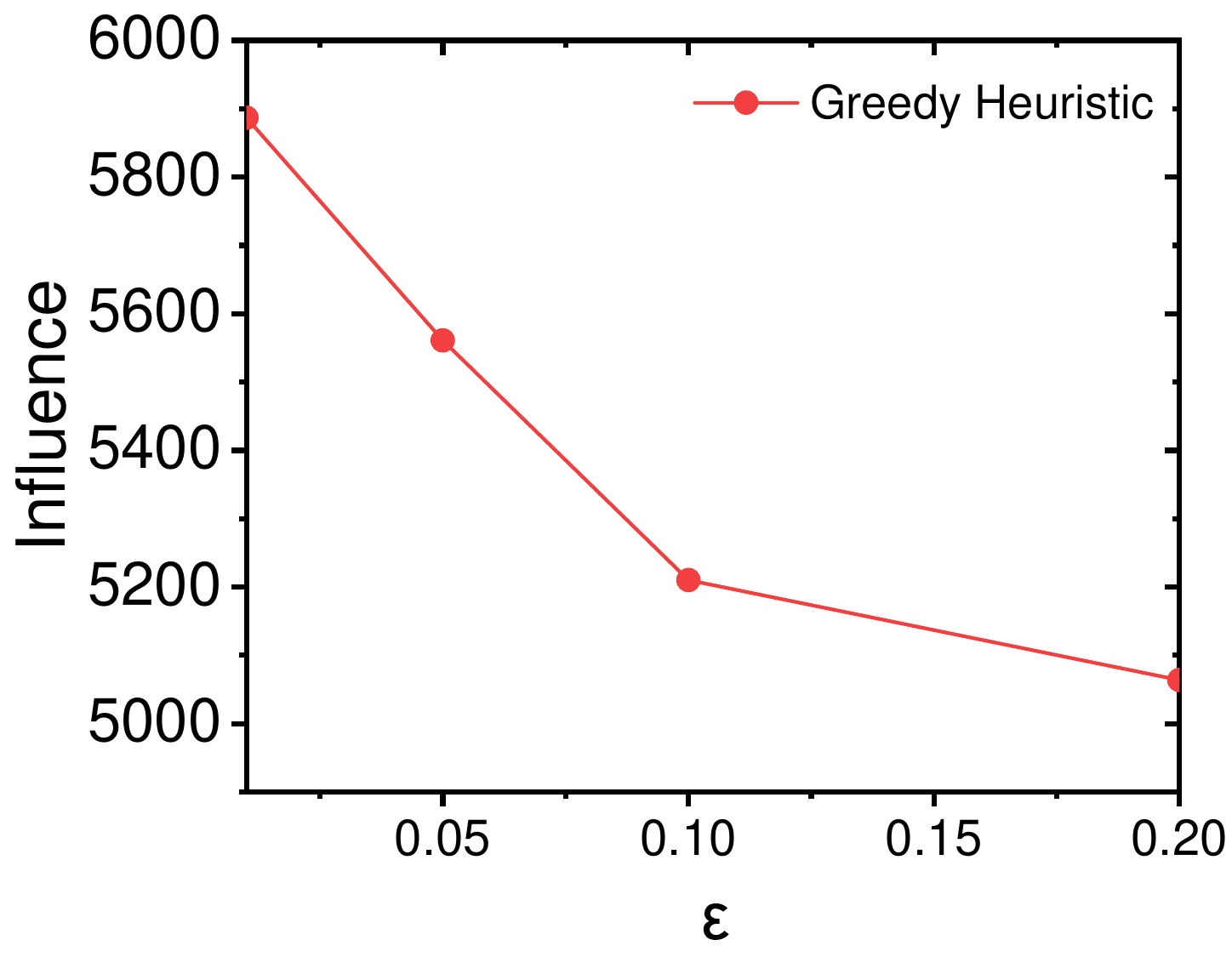} & \includegraphics[scale=0.12]{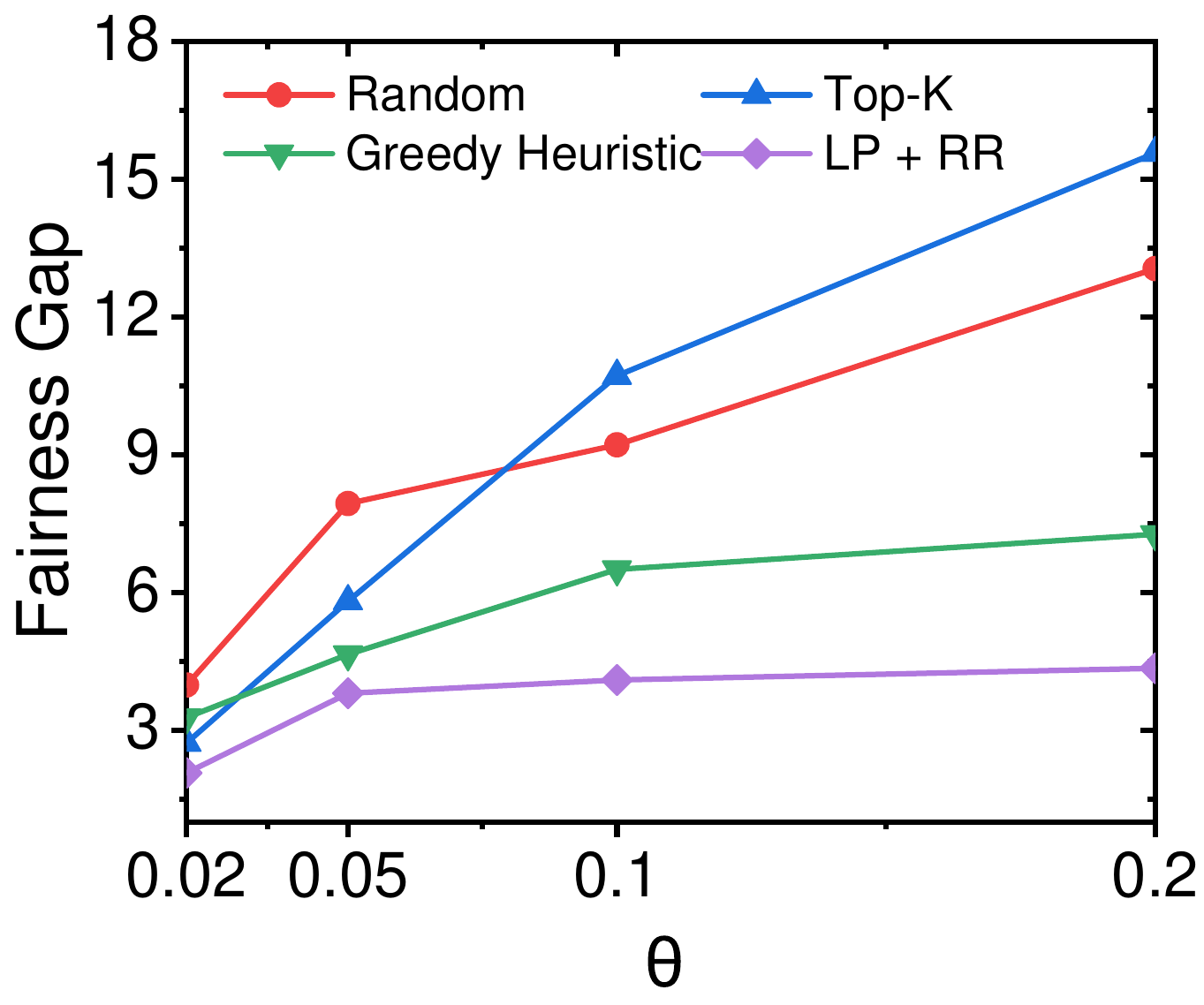} \\
\tiny{(e) $\epsilon =0.1$, $\theta = 0.05$} &  \tiny{(f) $\epsilon =0.1$, $\theta = 0.05$} & \tiny{(g) $\alpha = 100\%$, $\theta = 0.05$} & \tiny{(h) $\alpha = 100\%$, $\beta = 5\%$}  \\
\end{tabular}
\caption{ Algorithms Vs. Influence $(a,b,c,d)$, Varying $\alpha$ Vs. Fairness Gap (e), Time (f), Varying $\epsilon$ Vs. Influence (g), Varying $\theta$ Vs. Fairness Gap (h) with $|\mathcal{P}| = 20$, $\beta = 5\%$ for LA dataset}
\label{Fig:LA}
\vspace{-0.25 in}
\end{figure*}

\paragraph{\textit{Efficiency Test.}}
 Figure \ref{Fig:NYC}(f) and Figure \ref{Fig:LA}(f) show the efficiency of the proposed and baseline methods, and from this, we have three main observations. First, the run time of the `LP + RR' approach is higher compared to the `Greedy Heuristic' because solving the LP requires huge computational time. Further, randomized rounding, budget repair, and balance repair take additional computational time. The `Greedy Heuristic' takes less computational time because the sampling-based slot assignment reduces computational time. Second, with the increase of the demand-supply ratio $(\alpha)$, the run time for all the proposed and baseline methods increases because increasing $\alpha$ increases the demand for the slots of the products. Third, among baseline methods, `Top-$k$' takes a higher run time than `Random' because, before assigning slots to the products, the slots are sorted in descending order of their individual influence value.
\paragraph{\textit{Scalability Test.}} To show the scalability of the proposed approaches, we vary the number of products. We observe that the efficiency is very sensitive in `LP+RR' compared to the `Greedy Heuristic' when we vary the number of products from $5$ to $100$. It is observed that the runtime increase in `LP+RR' compared to `Greedy Heuristic' and baselines is almost $60\%$ to $80\%$ and $90\%$ to $95\%$.
\paragraph{\textit{Additional Discussions.}}
Now, we discuss the impacts of additional parameters, i.e., varying distance $(\lambda)$, Varying trajectory size, varying individual demand supply ratio $(\beta)$, etc. First, with the increase in the distance over which billboard slots can influence the trajectories, the influence of the products increases. Second, with the increase in trajectory size, the influence of all the proposed and baseline methods increases because one slot can influence a large number of trajectories. Third, with increasing individual demand supply ratio $\beta$, the demand for the number of slots for each product increases, and for this reason, the computational time for all proposed and baseline methods increases. In all our experiments, we set the value of $\alpha$, $\beta$, $\epsilon$, $\theta$, $\lambda$, and the trajectory size as $80\%$, $5\%$, $0.1$, $0.05$, $100$ meters, and $120k$ (NYC), respectively, as a default setting.
\vspace{-0.1 in}
\section{Concluding Remarks} \label{Sec:Conclusion}
In this paper, we have introduced and studied the Multi-Product Influence Maximization Problem for the Balanced Popularity Problem. We show that the problem is NP-hard to solve optimally. We formulate this problem as a linear programming problem and use linear programming relaxation with randomized rounding. Further, we propose a greedy-based heuristic with balance correction to solve this problem. All the methods have been analyzed to understand their time and space requirements. They have been implemented with real-world datasets and compared with baseline methods to show their effectiveness and efficiency.
However, the optimality has not been guaranteed. Now, our future work on this problem will remain concentrated on extending the methodologies for multiple advertisers.    
%
%
%
\bibliographystyle{splncs04}
\bibliography{Paper}

\end{document}